# News from CERN
# LHC Status and Strategy for Linear Colliders

Rolf-Dieter Heuer

CERN – Director-General
CH-1211 Geneva 23 – Switzerland

This paper presents the latest development at CERN, concentrating on the status of the LHC and the strategy for future linear colliders. The immediate plans include the exploitation of the LHC at its design luminosity and energy as well as upgrades to the LHC (luminosity and energy) and to its injectors. This may be complemented by a linear electron-positron collider, based on the technology being developed by the Compact Linear Collider and by the International Linear Collider and/or by a high-energy electron-proton collider. This contribution describes the various future directions, all of which have a unique value to add to experimental particle physics, and concludes by outlining key messages for the way forward.

## 1 The Large Hadron Collider Physics Programme

The Large Hadron Collider (LHC) [1] is primarily a proton-proton collider (see Figure 1) with a design centre-of-mass energy of 14 TeV and nominal luminosity of $10^{34}$ cm$^{-2}$s$^{-1}$, and will also be operated in heavy-ion mode. The high 40 MHz proton-proton collision rate and the tens of interactions per crossing result in an enormous challenge for the experiments and for the collection, storage and analysis of the data.

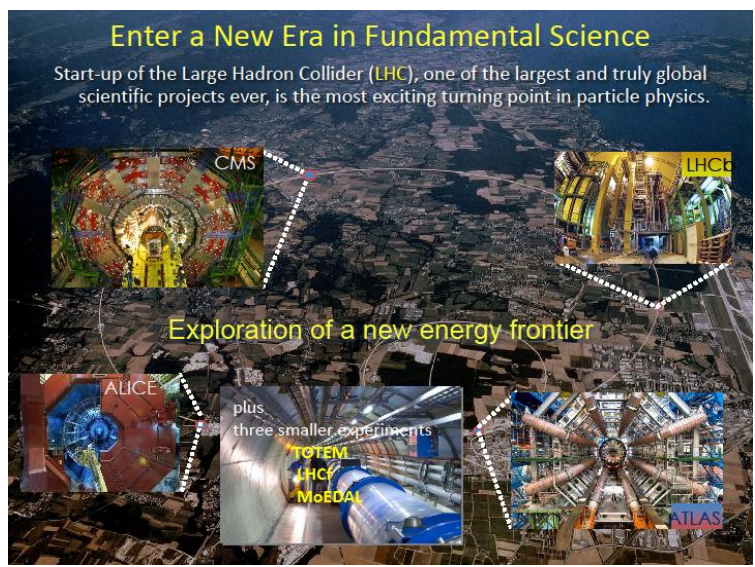

Figure 1: The LHC accelerator and the ALICE, ATLAS, CMS and LHCb experiments. There are also three smaller experiments - LHCf, MoEDAL and TOTEM.

By colliding unparalleled high-energy and high-intensity beams, the LHC is opening up previously unexplored territory at the TeV scale in great detail, allowing the experiments to probe deeper inside matter and providing further understanding of processes that occurred very early in the history of the Universe.

Of central importance to the LHC is the elucidation of the nature of electroweak symmetry breaking, for which the Higgs mechanism and the accompanying Higgs boson(s) are presumed to be responsible. In order to make significant inroads into the Standard Model Higgs Boson search, sizeable integrated luminosities of several $fb^{-1}$ are needed. However, even with 1 $fb^{-1}$ per experiment, discovery of the Standard Model Higgs Boson is still possible in mass regions beyond the lower limit of 114.4 GeV from direct searches at LEP2.

The LHC has provided the direct exclusion of a significant mass range unexplored until now. Exclusion limits have been placed on the Standard Model Higgs Boson by ATLAS and CMS. The mass region 146-466 GeV (except 232-256 GeV and 282-296 GeV) has been excluded by ATLAS at 95% CL while CMS has excluded the mass ranges 145-216 GeV, 226-288 GeV and 310-400 GeV.

At the initial LHC centre-of-mass energy of 7 TeV and with 1 $fb^{-1}$ per experiment, combining the results from ATLAS and CMS would provide a 3$\sigma$ sensitivity to a Standard Model Higgs Boson mass in the range 135 GeV to 475 GeV, and will exclude the Standard Model Higgs Boson between 120 GeV and 530 GeV at 95% CL. Combining the results from ATLAS and CMS at 7 TeV centre-of-mass energy and assuming about 10 $fb^{-1}$ per experiment would exclude at 95% CL the mass range from 600 GeV down to the LEP2 lower limit and would also provide a 3$\sigma$ sensitivity to a Standard Model Higgs Boson in the same mass range. Therefore, should the Standard Model Higgs Boson exist between the masses of 114 GeV to 600 GeV, it will either be discovered or ruled out by ATLAS and CMS until the end of 2012.

The reach for new physics at the LHC is considerable already at LHC start-up. In Supersymmetry (SUSY) theory, due to their high production cross-sections, squarks and gluinos can be produced in significant numbers even at modest luminosities. This would enable the LHC to start probing the nature of dark matter. No hint of SUSY particles has been observed so far, with the current lower limit being 1 TeV at 95% CL. in constrained SUSY models.

Moreover, the current lower limits for new heavy bosons Z' and W' are about 2.0 TeV, while those for excited quarks are about 3 TeV.

The LHCb experiment has been studying the physics in the B-meson sector. The experiment has studied $B_s$ oscillations and has measured a value of $\Delta m_s$ = 17.725 ±0.041± 0.026 $ps^{-1}$. LHCb and CMS have also been searching for the $B_s \rightarrow \mu\mu$ decay channel. This channel is predicted to be very rare in the Standard Model but has a large sensitivity to new physics, for example to SUSY. The current results are consistent with Standard Model predictions but given that the 95% CL is still 3.4 times the value in the Standard Model, there remains plenty of room for new physics to still appear.

The LHC will also provide information on the unification of forces, the number of space-time dimensions and on matter-antimatter asymmetry. With the heavy-ion collision mode, the LHC will probe the formation of the quark-gluon plasma at the origin of the Universe.

## 2  LHC Machine and Experiments Performance

The main LHC achievements for 2010 and 2011 can be summarized as follows:

- Excellent performance of the LHC machine for both proton-proton and Pb-ion runs. Beam operation availability was 65% on average. Peak instantaneous luminosities of $2 \times 10^{32}$ cm$^{-2}$ s$^{-1}$ were attained for proton-proton collisions, which were a factor of two above the 2010 goal and which resulted in almost 50 pb$^{-1}$ of integrated luminosity delivered to the experiments. Following a short 4-day switch-over to Pb-ion beams, peak luminosities of $3 \times 10^{25}$ cm$^{-2}$ s$^{-1}$ were attained for Pb-Pb collisions with almost 10 μb$^{-1}$ of integrated luminosity delivered to the experiments. This is a great achievement for the first full year of LHC operation.

  The excellent performance of the LHC machine was also reported for 2011 until now. An integrated luminosity of more than 3.5 fb$^{-1}$ has been delivered to each of the ATLAS and CMS experiments.

- All experiments took data of excellent quality and with high efficiency and they have been coping well with multiple interactions per crossing. The physics analyses re-measured the science of the Standard Model of Particle Physics, in many instances superseding limits set at the Tevatron while taking the LHC's first steps into new territory. As a result, a plethora of physics papers were published and conference presentations were made by the LHC experiments.

- The performance of the Worldwide LHC Computing Grid (WLCG) was also outstanding, exceeding the design bandwidth and allowing a very fast reconstruction and analysis of the data.

## 3  The LHC Consolidation and Upgrades

The coming years will lay the foundation for the next decades of high-energy physics at the LHC. The LHC research programme until around 2030 is determined by the full exploitation of its physics potential, consisting of the design luminosity and the high-luminosity upgrade (HL-LHC). Together with superconducting higher-field magnets for a higher-energy proton collider (HE-LHC), if necessitated by the physics, these initiatives will position CERN as the laboratory at the energy frontier.

### 3.1 LHC Consolidation

The LHC proton-proton and Pb-ion operations periods in 2011-2012 will be followed by a long shutdown in 2013 and extending well into 2014 with the following objectives:

- To repair and consolidate the inter-magnet copper-stabilizers (splices) to allow for safe operation up to 7 TeV/beam for the lifetime of the LHC.

- In the shadow of the inter-magnet copper-stabilizer work, the installation of the pressure rupture disks (DN200) will be completed and around 20 magnets which are known to have problems for high energy will be repaired or replaced. In addition, PS and SPS consolidation and upgrade work will be carried out.

- During this shutdown, the collimation system will also be upgraded at Point 3.

- The experiments will use the shutdown to implement a programme of consolidation, improvements and upgrades.

### 3.2 High Luminosity LHC

The strategy for the LHC for the coming years is the following:

- Exploitation of the physics potential of the LHC up to design conditions in the light of running experience and by optimizing the schedule for physics.

- Preparation of the LHC for a long operational lifetime through appropriate modifications and consolidation to the machine and detectors and through the build-up of an adequate spares inventory.

- In the years until 2018, the LHC will be operated towards 7 TeV/beam with increased intensities and luminosities.

- In 2018, a long shutdown is scheduled to connect LINAC4 [2], to complete the PS Booster energy upgrade, to finalize the collimation system enhancement and to install LHC detector improvements. After this shutdown, a further period of three years of LHC operation at 7 TeV/beam and at least the design luminosity is planned (with short technical stops around the end of each year).

- The ambitious longer-term plans include a total integrated luminosity of the order of 3000 $fb^{-1}$ (recorded) by the end of the life of the LHC. The HL-LHC implies an annual luminosity of about 250-300 $fb^{-1}$ in the second decade of running. The HL-LHC upgrade is also required to implement modifications to elements in the insertion regions of the machine whose performance will have deteriorated due to radiation effects, such as the inner triplet quadrupole magnets. The HL-LHC upgrade is scheduled for the 2022 long shutdown.

- LHC detector R&D and upgrades to make optimal use of the LHC luminosity.

This strategy is also driven by the necessity to bring the LHC injector chain and the technical and general infrastructure up to the high standards required for a world laboratory in order to ensure reliable operation of the CERN complex.

### 3.3 Higher Energy LHC

Increasing the beam energy beyond the 7 TeV nominal energy of the LHC can be obtained by raising the mean field in the dipole magnets. The main parameter of the HE-LHC machine is the 16.5 TeV beam energy, and is based on the hypothesis of substituting all the present dipoles with new, more powerful ones, capable of operating at 20 T with beam. The success of the HE-LHC study depends critically on the success of the magnet R&D to reach dipole fields around 20 T in a useful bore. Despite the variety of superconducting materials and the continuous new discoveries, the practical superconductors (*i.e.* with good physics characteristics, good workability and suitability to cabling) are limited. The candidate materials are NbTi, $Nb_3Sn$, $Nb_3Al$ and high-temperature superconductor (HTS). Additional open issues that require continuing R&D include high-gradient quadrupole magnets for the arcs and the interaction regions; fast-cycling superconducting magnets for a 1-TeV injector; emittance control in the regime of strong synchrotron damping; cryogenic handling of the synchrotron radiation heat load; and a study of the dynamic vacuum. The HE-LHC studies have started at CERN within an international collaboration and a first estimate provides a timeline of between 2030-2033 to start the realization of the HE-LHC machine project.

## 4 Colliders at the Energy Frontier beyond the LHC

Great opportunities are in store at the TeV scale and a fuller understanding of Nature will come about through a clearer insight at this energy level. The LHC will provide a first indication of any new physics at energies up to several TeV. First results from the LHC will be decisive in indicating the direction that particle physics will take in the future. Many of the open questions left by the LHC and its upgrades may be addressed best by an electron-positron collider, based on technology developed by the Compact Linear Collider (CLIC) [3] and International Linear Collider (ILC) [4] collaborations. Options for a high-energy electron-proton collider (LHeC) [5] are also being considered. As in the past, there is a synergy between collider types that can be used to advantage. The discovery of the Standard Model over the past few decades has advanced through the synergy of hadron-hadron (e.g. SPS and the Tevatron), lepton-hadron (HERA) and lepton-lepton colliders (e.g. LEP and SLC). Such synergies should be continued in the future and thus a strategy has been developed along these lines. The above effort on accelerators should advance in parallel with the necessary detector R&D.

## 4.1 Linear Electron-Positron Collider

### 4.1.1 The Compact Linear Collider (CLIC)

The conceptual lay-out of the CLIC is shown in Figure 2. The collider is based on using lower-energy electron beams to drive high-energy electron and positron beams. The fundamental principle is that of a conventional AC transformer. The lower-energy drive beam serves as a radiofrequency source that accelerates the high-energy main beam with a high accelerating gradient. The nominal centre-of-mass energy is up to 3 TeV, the nominal luminosity exceeds $10^{34}$ cm$^{-2}$ s$^{-1}$, the main linear accelerator frequency is 12 GHz, the accelerating gradient is 100 MeV/m and the total length of the main linear accelerators is up to 48.3 km.

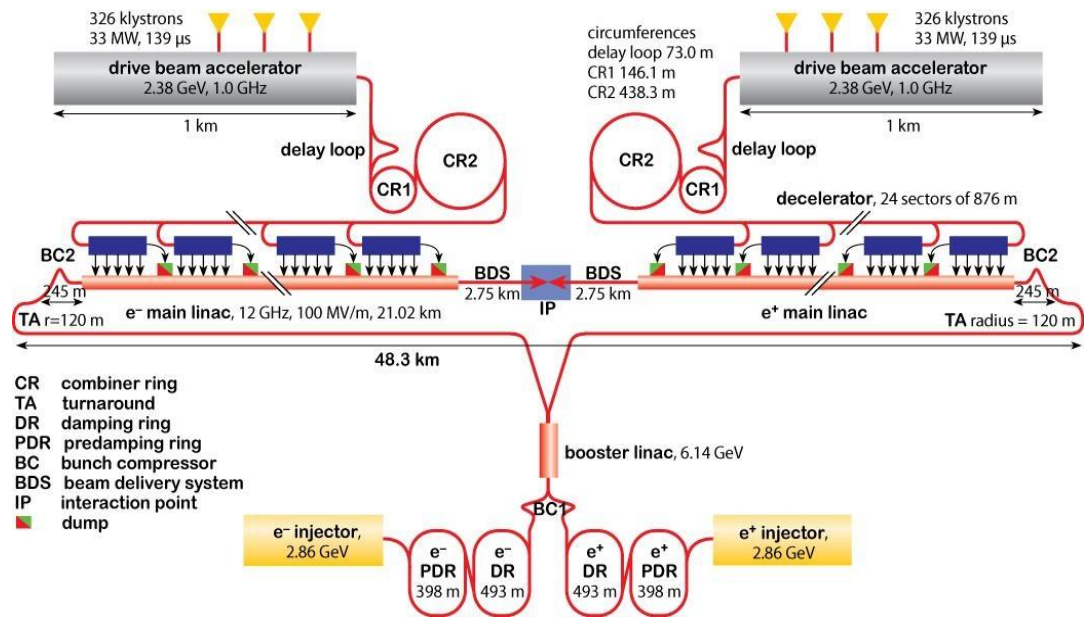

Figure 2: The CLIC general lay-out.

The CLIC project is in its R&D phase. In particular, the target accelerating gradient is considerably high and requires very aggressive performance from the accelerating structures. The nominal CLIC accelerating gradient has been exceeded in an unloaded structure with a very low breakdown probability of less than $3 \times 10^{-7}$ per metre for the nominal pulse length after conditioning of the radiofrequency cavities for 1200 hours.

The mandate of the CLIC team is to demonstrate the feasibility of the CLIC concept, to be published by 2012 in a Conceptual Design Report. If this effort is successful, and if the new physics revealed by the LHC warrants, the next phase of R&D on engineering and cost issues will be launched. This would serve as the basis for a Technical Design Report and a request for project approval.

*4.1.2 The International Linear Collider (ILC)*

The ILC, shown in Figure 3, is an option for a linear electron-positron collider at lower energies than CLIC and is based on a more conventional design for acceleration using superconducting standing wave cavities with a nominal accelerating field of 31.5 MeV/m and a total length of 31 km at 500 GeV centre-of-mass energy. A two-stage technical design phase during 2010-2012 is presently underway, culminating in a Technical Design Report. A major contribution from Europe and from DESY to the ILC Global Design Effort is the European X-ray Laser Project XFEL at DESY. The purpose of the facility is to generate extremely brilliant and ultra-short pulses of spatially-coherent X-rays. The electron energy is brought up to 20 GeV through a superconducting linear accelerator, of length one-tenth that of the ILC superconducting linear accelerator, and conveyed to long undulators where the X-rays are generated and delivered to the experimental stations. Construction of the XFEL is underway and operation is planned to start in 2015.

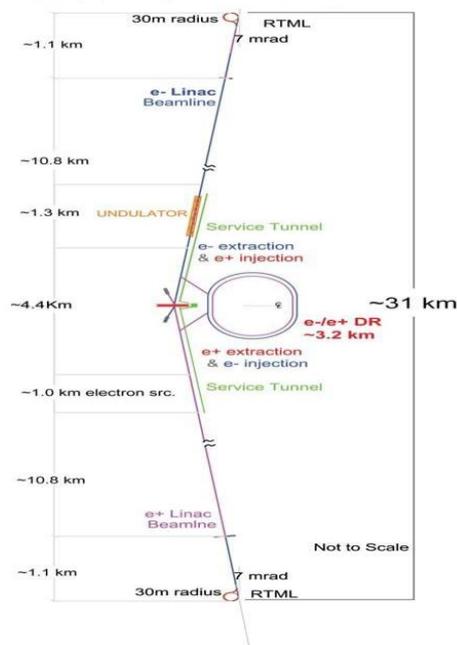

Figure 3: A schematic lay-out of the ILC.

*4.1.3 Towards a Single Linear Collider Project*

The strategy to address key issues common to both linear colliders involves close collaboration between ILC and CLIC. Recent progress has been encouraging in this respect and a first common meeting between ILC and CLIC was held in February 2008. Such close collaboration would facilitate the exchange of concepts and R&D work between the ILC and CLIC in the future. In the meantime, items to be addressed as a matter of high priority for both linear collider projects include the construction costs, power consumption and value engineering.

*4.1.4 Detector Challenges*

R&D on key components of the detector for a linear collider is mandatory and is also well underway. High-precision measurements demand a new approach to the reconstruction. Particle flow, namely reconstruction of all particles, is thus proposed and requires unprecedented granularity in three dimensions of the detection channels.

## 4.2 Lepton-Hadron Collider

The option of a high-energy electron-proton collider - the Large Hadron Electron Collider (LHeC) - is being considered for the high-precision study of QCD and of high-density matter at the energy frontier. The LHeC design consists of an electron beam of 60 GeV (to possibly 140 GeV) colliding with protons of 7 TeV energy from the LHC. Assuming an electron-proton design luminosity of about $10^{33}$ cm$^{-2}$ s$^{-1}$, the LHeC could exceed the corresponding HERA values of integrated luminosity by two orders of magnitude and the kinematic range by a factor of 20 in the four-momentum squared $Q^2$ and in the inverse Bjorken *x*. The projected physics reach of the LHeC is shown in Figure 4. Both a ring-ring option - consisting of a new ring in the LHC tunnel with bypasses around the LHC experiments - and a linac-ring option - based on a re-circulating linac with energy recovery or on a straight linac - are being considered. The LHeC Conceptual Design Report will be completed by the end of 2011.

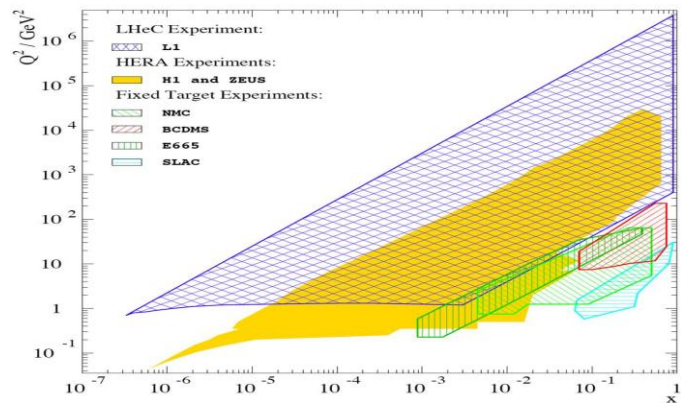

Figure 4: The projected physics reach of LHeC.

## 4.3 Generic Accelerator R&D

In addition to continuing focused R&D on the above collider projects, R&D should also be supported for generic accelerator studies. For example, R&D for high-power proton sources, such as the high-power superconducting proton linac (HP-SPL), in line with European participation in neutrino physics, should continue. Moreover, studies for novel acceleration techniques based on plasma acceleration is mandatory, as such machines could potentially provide the high acceleration gradients to reach the high energies required by the physics at future colliders.

# 5 The European Strategy for Particle Physics

European particle physics is founded on strong national institutes, universities and laboratories, working in conjunction with CERN. The increased globalization, concentration and scale of particle physics require a well-coordinated European strategy. This process started with the establishment of the CERN Council Strategy Group, which organized an open symposium in Orsay in early 2006, a final workshop in Zeuthen in May 2006 and with the strategy document being signed unanimously by Council in July 2006 in Lisbon [6]. CERN considers that experiments at the high-energy frontier to be the premier physics priority for the coming years. This direction for future colliders at CERN follows the priorities set in 2006 by the CERN Council Strategy Group. The years 2010 and 2011 are seeing the start of the LHC physics exploitation leading to important input for the update of the European strategy for particle physics planned for 2012-2013.

# 6 Key Messages

Particle physics will need to adapt to the evolving situation. Facilities for high-energy physics (as for other branches of science) are becoming larger and more expensive. Funding for the field is not increasing and the timescale for projects is becoming longer, both factors resulting in fewer facilities being realized. Moreover, laboratories are changing their missions.

All this leads to the need for more co-ordination and more collaboration on a global scale. Expertise in particle physics needs to be maintained in all regions, ensuring the long-term stability and support through-out. It would be necessary to engage all countries with particle physics communities and to integrate the communities in the developing countries. The funding agencies should in their turn provide a global view and synergies between various domains of research, such as particle physics and astroparticle physics, should be encouraged. Particle physics has entered a new and exciting era. The start-up of the LHC allows particle physics experiments at the highest collision energies. The expectations from the LHC are great, as it would provide revolutionary advances in the understanding of the microcosm and a fundamental change to our view of the early Universe. Due to the location of the LHC, CERN is in a unique position to contribute to further understanding of the microcosm in the long term.

Results from the LHC will guide the way in particle physics for many years. It is expected that the period of decision-making concerning the energy frontier will be in the next few years. Particle physics is now in an exciting period of accelerator planning, design, construction and running and would need intensified efforts in R&D and technical design work to enable the decisions for the future course and global collaboration coupled with stability of support over long time scales.

The particle physics community needs to define now the most appropriate organizational form and needs to be open and inventive in doing so, and it should be a dialogue between the scientists, funding agencies and politicians. It is mandatory to have accelerator laboratories in all regions as partners in accelerator development, construction, commissioning and exploitation. Furthermore, planning and execution of high-energy physics projects today require world-wide partnerships for global, regional and national projects, namely for the whole particle physics programme. The exciting times ahead should be used to advantage to establish such partnerships.

## 7 Opening the Door

CERN Council opened the door to greater integration in particle physics when it recently unanimously adopted the recommendations to examine the role of CERN in the light of increasing globalization in particle physics. The key points agreed by Council include a) all states shall be eligible for CERN Membership, irrespective of their geographical location; b) a new Associate Membership status is to be introduced to allow non-Member States to establish or intensify their institutional links with CERN; and c) the participation of CERN in global projects is to be enabled wherever they are sited.

Several countries are now in the process of greater integration with CERN. Romania is a Candidate to Accession to CERN Membership; negotiations with Israel for Associate Membership as a pre-stage to full Membership have been concluded; and negotiations with Cyprus, Serbia, Slovenia and Turkey for Associate Membership have started. A number of other countries have also expressed interest in Associate Membership.

## 8 Conclusions

In this paper, we have reported on the latest developments at CERN, such as the status and future plans of the LHC machine and experiments and also presented the strategy for future linear colliders. In the coming years, the priorities are the full exploitation of the LHC together with preparations for a luminosity upgrade. Moreover, studies for increasing the beam energy of the LHC are underway. It will be necessary to keep under review the physics drivers for future proton accelerator options and it will be necessary to compare the physics opportunities offered by proton colliders with those available at a linear electron-positron collider. The R&D associated with future colliders needs to continue in parallel.

## 9 Acknowledgements

I would like to thank the organizers for the invitation to make this contribution and for the excellent organization of the very interesting conference, which also provided the opportunity to present the latest developments at CERN. Many thanks go to Emmanuel Tsesmelis for his assistance in preparing this contribution.